\documentclass[12pt]{article}
\usepackage[usenames,dvipsnames]{xcolor}  %taken on 13/7/12 from %http://en.wikibooks.org/wiki/LaTeX/Colors
\usepackage{amsfonts,amssymb,amsmath,amsthm}
\usepackage{url}
\usepackage{authblk}
\usepackage[toc,page]{appendix}
\usepackage{epsfig}
\usepackage{bm} %Makes bold of math expressions, including spacings. e.g. $\bm{\alpha \ne \beta}$
\normalbaselines %\renewcommand{\baselinestretch}{1.565} % use, if double-spacing is required
\newcommand{\nc}{\newcommand}  %\nc must be defined {\em before} it is used.
\def\comm#1{}   % this definition has to be at the beginning
\def\edd{\end{document}}
\def\edda{\end{appendices}  \end{document}}
\nc\bes{\begin{subequations}}
\nc\ees{\end{subequations}}\nc\wbar{\widebar{subequations}}
\nc{\cc}{commutation relations~}
\nc{\floor}[1]{\lfloor{#1}\rfloor}
\newcommand{\semiplus}{+ \!\!\!\!\!\! \supset}

\newcommand{\sqt}{\sqrt{2}}
\newcommand{\x}{\times}
\newcommand{\ba}{\begin{array}}       % to start a matrix
\newcommand{\ea}{\end{array}}
\newcommand{\s}{\sigma}
\newcommand{\crn}{\nn \\[2mm]}
\newcommand{\hhline}{\hline \hline}  \nc{\dg}{\dagger}

\newcommand{\om}{\omega}
\newcommand{\so}{{\mathfrak {so}}} % so(N) algebra

\nc\spr[1]{\langle #1 \rangle}    %\def\sp #1 {\langle #1 \rangle}
\nc\hnomial[2]{\left[ #1 \atop #1 \right]}
\nc\hc[1]{\left[ #1 \right]}
\nc\ket [1]{|#1 \rangle}  %\def\ket#1{|#1 \rangle}
\nc\bra [1]{\langle #1|}
\nc{\g}{{\mathfrak {g} }}   % g Lie algebra
\def\h{{\mathfrak {h} }}   % Heisenberg algebra
\def\su{{\mathfrak {su}}} % su(N) algebra
\def\gl{{\mathfrak {gl}}} % gl(N) algebra
\def\sl{{\mathfrak {sl}}} % sl(N) algebra
\def\half{\frac 1 2} 

\nc{\ph}{\varphi} %newlabel
\nc{\al}{\alpha} \nc{\ga}{\gamma} \nc{\la}{\lambda}
\nc{\tx}{\tilde{x}} \nc{\tp}{\tilde{p}}
\nc{\eps}{\epsilon}

\newcommand{\tr}{\mbox{Tr}\,}   % trace
\def\pd#1#2{\frac{\partial #1}{\partial #2} }  % newcommand pd already used

\newcommand{\lb}[1]{\label{#1}}
\newcommand{\rf}[1]{(\ref{#1})}
\newcommand{\be}{\begin{equation}}

\newcommand{\ee}{\end{equation}}
\newcommand{\el}[1]{\label{#1}\end{equation}}
\newcommand{\erl}[1]{\label{#1}\end{eqnarray}}
\newcommand{\br}{\begin{eqnarray}}
\newcommand{\er}{\end{eqnarray}}

\newcommand{\Ra}{\Rightarrow }
\newcommand{\nn}{\nonumber}

\newcommand{\for}{\qquad \mbox{ for}\quad}
\newcommand{\aand}{\quad \text{ and}\quad}

\newcommand{\where}{\quad \mbox{ where}\quad}

\newcommand{\ie}{{\em i.e.~}}
\newcommand{\sbs}{\subsection}
\newcommand{\sbb}{\subsubsection}
\def\R{{\mathbb R}} % reelle Zahlen
\def\Z{{\mathbb Z}} % ganze Zahlen
\def\hc#1#2{\left[\! \! \begin{array}{c}#1 \\ #2\end{array} \! \! \right]}  % Hermite coefficients
\nc{\hcn}[1]{\left[#1\right]}  % new Hermite coefficients a \atop b
\nc\ah{\hat{a}}\nc\hx{\hat{x}}\nc\hp{\hat{p}}
\nc\pt{\partial_t}
\nc\px{\partial_x}
\nc\prp{\partial_p}
\nc\py{\partial_y}
\nc\pu{\partial_u}
\newcommand{\gen}[1]{\partial_{#1}}

\newcommand{\pdx}{~\partial_x}
\newcommand{\pdp}{~\partial_p}
\newcommand{\pdt}{~\partial_t}
\newcommand{\dx}{\pd{}{x}}

\newcommand{\dt}{\pd{}{t}}

\newtheorem{prop}{Proposition}

\newtheorem{cor}{Corollary}
\newtheorem{rmk}{Remark}

\numberwithin{equation}{section}

\begin{document}

\title{Symmetries of the pseudo-diffusion equation, and its unconventional 2-sided kernel}

%to whom the original three partners expressed their gratitude in the rough draft for his hospitality

\author[1]{Jamil \ Daboul \thanks{daboul@bgu.ac.il}}
\author[2]{Faruk G\"ung\"or \thanks{gungorf@itu.edu.tr}}
\author[3]{Dongsheng\ Liu \thanks{Work on the present paper was started in 1998 by a collaboration of J. Daboul, D. Liu and D. McAnally. A draft was produced by correspondence in 2000 and remained essentially dormant till the April 2015. Finally, the paper was
expanded and completed by J. Daboul and F. G\"ung\"or. Subsequently, Daboul tried without success to locate his old partners via the Internet. Finally, Daboul was informed by Prof. R. Delbourgo that Liu is working somewhere in Melbourne and that D. McAnally tragically died in 2006.}}
 %D.Liu@utas.edu.au}}
\author[4]{David~S~McAnally} %\thanks{dsm@maths.uq.edu.au}}
\affil[1]{Physics Department, Ben Gurion University of the Negev, 84105 Beer Sheva, Israel}
\affil[2]{Department of Mathematics, Faculty of Science and Letters, Istanbul Technical University, 34469 Istanbul, Turkey}
\affil[3]{School of Mathematics and Physics, University of Tasmania\\
 GPO Box 252C, Hobart, Australia 7001}
\affil[4]{University of Melbourne, Department of Mathematics and Statistics, Melbourne, Australia 3052}

\date{\today}
\maketitle

\begin{abstract}
We determine by two related methods the invariance algebra $\g$ of the \emph{`pseudo-diffusion equation'} (PSDE)
$$
L~Q \equiv \left[\frac {\partial}{\partial t}
-\frac 1 4 \left(\frac {\partial^2}{\partial x^2} -\frac 1 {t^2}
\frac {\partial^2}{\partial p^2}\right)\right]~Q(x,p,t)=0,
$$
which describes the behavior of the $Q$ functions in the $(x,p)$-phase space as a function of a squeeze
parameter $y$, where $t=e^{2y}$.

The  algebra turns out to be isomorphic to that of its constant coefficient version. Relying on this isomorphism we construct a local point
transformation which maps the factor $t^{-2}$ to 1.

We show that any generalized version $u_t-u_{xx}+ b(t) u_{yy}=0$ of PSDE has a smaller symmetry algebra than $\g$,
except for $b(t)$ equals to a constant or it is proportional to $t^{-2}$.

We apply the group elements $G_i(\ga) := \exp[\ga A_i]$ and obtain new solutions of
the PSDE from simple ones, and interpret the physics of some of the results.

We make use of the `factorization property' of the PSDE to construct its \textit{`2-sided kernel'}, because it has to depend on two times, $t_0 < t < t_1$.

We include a detailed discussion of the identification of the Lie algebraic structure of the symmetry
algebra $\g$, and its contraction from  $\su(1,1)\oplus\so(3,1)$.

\end{abstract}

\section{Introduction}

The following partial differential equation has been called the {\em pseudo-diffusion equation} (PSDE) by
\cite{md, mm, dab, dmm}
\be
L(x,p,t)\, Q(x,p,t) \equiv
\left(\frac{\partial}{\partial t}-\frac{1}{4}\frac{\partial^2}{\partial x^2}
     +\frac{1}{4t^2}\pd{^2}{p^2}\right) Q(x,p,t)=0,
\el{psd}
It describes a diffusive process in $x$, and an \emph{`infusive'} one in $p$ for all $t$ since the second
derivative in $p$ has the opposite sign. Intuitively, one may think of $t$ as time, and of $x$ and $p$ as
two Cartesian coordinates.

Eq.\rf{psd} was originally derived \cite[Sec. 2.1]{dmm} in the form,
\be
\frac{\partial}{\partial y} \hat{\Pi}(x,p\,;y) = \frac 1 2 \left(
e^{2y}\frac{\partial^2}{\partial x^2} - e^{-2y}\pd{^2}{p^2}\right) \hat{\Pi}(x,p\,;y)
\el{psd2}
where $\hat{\Pi}(x,p \,;y)$ is the \emph{Husimi projection operator} \cite{husimi}:
\be
\hat{\Pi}(x,p\,;y):=\ket{x,p \,;y}\bra{x,p \,;y}\,,
\el{proj}
where $\ket{x,p\,;y}$ are \emph{squeezed coherent states}, defined in Eq.\rf{pqy2}. The $y$ is related to the complex \emph{`squeezing parameter'} $\zeta = y\ e^{i\ph}$ (see \rf{S1})\,;
For clarity and establishing the notation, we shall review in Sec. 2 the relevant concepts and definitions needed to understand equations \rf{psd} and \rf{psd2}. For a historical review, see Dodonov \cite{dodonov}.

For now we mention that the projection operator $\hat{\Pi}(x,p\,;y)$ in \rf{proj} is used to map \emph{quantum-mechanical operators}
$\hat O$ (such as density operators $\hat \rho$),  which are usually defined in terms of the position coordinate $\hat O(x)$ or in terms of the momentum variable $\tilde O(p)$ to semi-classical distribution \textit{functions} $\tilde {Q}(\hat O:x,p\,;y)$ in $(x,p)$-phase space, as follows:
\be
Q(\hat O:x,p\,;t=e^{2y}) \equiv \tilde{Q}(\hat{O}:x,p\,;y):= \tr[\hat O\, \hat{\Pi}(x,p \,; y)]
= \spr{x,p\,; y|\hat O|x,p\,; y}.
\el{Q}
Thus, Eq.\rf{psd} describes how $Q(\hat O:x,p\,; t)$ is redistributed continuously in the $(x,p)$ phase
space as we change the \textit{real} squeezing parameter $y$, where $t=e^{2y}$.

Equation \rf{psd} is interesting mathematically in its own right,
regardless of its application to squeezing. We shall therefore first study its symmetry in Section 3 and
identify its symmetry algebra $\g$. In particular, we shall show that the $1/t^2$ factor in \rf{psd}
is crucial for the PSDE  \rf{psd} to have the maximum symmetry algebra of nine generators.
These generators will be derived by two methods:
\begin{enumerate}
	\item A simplified procedure which is applicable for \emph{linear} partial differential equations.		This method	will be explained in some details in Appendix A, for the benefit of non-experts.

\item The nine generators $X_i$, $i=1,\ldots,9$ given in \rf{Xi} will be obtained by \emph{the prolongation method} due to
S. Lie (see for example \cite{olver}). This will be done in Sec. 3.
\end{enumerate}
In Sec. 4 we derive a (discrete) symmetry of the PSDE equation under the exchange of the variables $x$ and $p$ and also discuss its
`factorization'. We apply these results to construct an unconventional 2-sided kernel, defined by initial conditions which
depend on two different time limits in order to produce the usual product of two delta functions $\delta(x-x_0) \delta(p-p_0)$ (see \rf{kxp}).

In Sec. 5 we apply the group elements $G_i(\ga) := \exp[\ga A_i]$ in \rf{groups} to obtain complicated solutions of
the PSDE \rf{psd} from simpler ones. We shall also interpret the physics of
some of the results.
In Sec. 6 we study several applications of the symmetries.

Finally, in Appendix B we present a detailed explanation of the algebraic structure of the symmetry algebra $\g$ and its interesting derivation of $\g$ from the contraction from the
Lie algebra $\su(1,1)\oplus \so(3,1)$.

%%%%%%%%%%%%%%%%%%%%%%%%%%%%%%%%%%%%%%%%%%%%%%%%%%%%%%%%%%%
\section{A review of coherent states and squeezing formalism}
%%%%%%%%%%%%%%%%%%%%%%%%%%%%%%%%%%%%%%%%%%%%%%%%%%%%%%%%%%%

In the present paper we restrict our definitions of coherent states and squeezing to solutions of the one-dimensional harmonic oscillator.

\subsection{Displacement operator $D(\al)$, coherent and squeezed states
and Weyl representation}

The {\em displacement or Weyl operator} $D(\al)$ is a unitary operator,
defined by
\be
D(\al) := \exp\ [\alpha \ah^\dg -\alpha^* \ah] =
\exp\ [i (p \hx -x \hp)] \equiv D(x,p),
\el{dis}
where $\hp=-i\hbar\, \gen x $ is the momentum operator and $\ah^\dg$ is the creation operator of the harmonic oscillator
\be
\ah^\dg \equiv \frac {\hx - i \hp} {\sqrt{2}} = \frac 1 {\sqrt{2}} \left(\frac x {x_0}- x_0\, \dx \right), \quad
(x_0=\sqrt{\hbar/m\om}\,), \aand \alpha \equiv \frac {x+i p}{\sqt}.
\el{aa}
Glauber in 1963 \cite{glauber} defined a {\em coherent state} $\ket{\al}$ by applying $D(\al)$ on $\ket{0}$, the ground-state of the oscillator. One obtains
\bes \lb{coh} \begin{align} %\end{align} \ees
\ket{x,p} &\equiv \ket{\al} := D(\al) \ket{0} = \exp\ [-|\alpha|^2/2 +\alpha \ah^\dag]\, \ket{0} \lb{coh1} \\[2mm]
          &= \exp\ [-|\alpha|^2/2]~ \sum_{n=0}^\infty \frac{\alpha^n}{\sqrt{n!}} ~\ket{n} ,
\end{align}
\ees
where the \emph{number states} $\ket{n}$ are eigenstates of the Hamiltonian $H = \om (\ah^\dag \ah + \frac 1 2)$ of the oscillator.

The following unitary operator is called \emph{squeezing operator}:
\bes \lb{S}
\begin{align}
S(\zeta) &:= \exp\ [\half(\zeta \ah^{\dg 2} -\zeta^* \ah^2], \where \zeta = y e^{i\ph},
\quad -\infty <y < \infty \lb{S1} \\
&\Ra   \exp\ [-i \frac y 2(\hx\,\hp +\hp\,\hx)], \for \ph=0. \lb{S2}
\end{align}
\ees
Stoler \cite{stoler} defines the squeezed coherent states by
\be
\ket{ \zeta\,;\al}:= S( \zeta) D(\al)\ket{0},
\el{stoler}
but in \cite{dmm} (and also in \cite{kim}, for example) they are defined in the opposite order, as follows
\bes \lb{pqy}
\begin{align}
\ket{\al\,; \zeta} &:= D(\al) S(\zeta) \ket{0}\,, \quad  \text{so that for real $\zeta$, we obtain} \lb{pqy1} \\[2mm]
\ket{x,p\,;y} &:= D(x,p) S(y) \ket{0} = \exp\ [i (p \hx-x \hp)] \exp\ [-i \frac y 2(\hx\,\hp +\hp\,\hx)] \ket{0}. \lb{pqy2}
\end{align}
\ees
Both definitions in \rf{stoler} and \rf{pqy} yield equivalent results. We shall adopt the
definitions in \rf{pqy}.

The coherent states $\ket{x,p}=\ket{x,p \,; y=0}$ and the squeezed states $\ket{x,p \,; y\ne 0}$ with
\emph{real} $\zeta$ are \emph{minimal wave packets}, \ie they satisfy the minimum Heisenberg uncertainty
relation $(\Delta x)^2 \cdot (\Delta p)^2 = \hbar^2/4$.  The coherent states $\ket{x,p}$ satisfy in addition the equality $(\Delta x)^2 = (\Delta p)^2$,
while the squeezed states have unequal dispersions, $(\Delta x)^2 \ne (\Delta p)^2$ \cite[Eqs.(18)] {stoler}.

By manipulating the projection operator \rf{proj} the PSDE equation was derived first in the form \rf{psd2}
\cite[Eq.(26)] {dmm}, where $\ket{x,p \,; y}$ is defined in \rf{pqy2}.

\section{The symmetry algebra}

In this section we look for infinitesimal symmetry generators $A$ for
the PSDE \rf{psd}.
For our purposes, we call a generator $A$ of the form
\be
A =\alpha(x,p,t)\partial_t+\beta(x,p,t)\partial_x +\gamma(x,p,t)
\partial_p +\eta(x,p,t),
\el{a}
an \emph{infinitesimal symmetry generator}, if for every
solution $u(x,p,t)$ of the PDE \rf{psd}, $A\,u(x,p,t)$ is also be a solution of \rf{psd}.

We now show that \emph{if $A$ satisfies the commutation condition
\be
[L, A]=\xi(x,p,t) L,
\el{rq}
then A is an infinitesimal symmetry generator.}

Let $u(x,p,t)$ be a solution of \rf{psd}, \ie $L\,u=0$. Then,
$$
L\cdot A u = (L A)u = ([L,A] + A L) u = (\xi + A) L u = 0,
$$
which means that $A\,u$ is also a solution. Furthermore, $A^j u$, $j=2,3,\dots$ and $\exp (\ga A)u$
are all solutions if $u$ is one solution.

In appendix A we apply the condition \rf{rq} to obtain a full set of
symmetry generators of PSDE \rf{psd}.

\subsection{The ansatz for the infinitesimal symmetry operators}
As Eq.\rf{psd} is linear two different approaches can be used to find the symmetry algebra.
One is the  \emph{Lie prolongation method}, another is the method of linear operators usually adopted
by physicists. We shall use both methods to give an idea how they are related. Here we present the first
approach and refer to Appendix \ref{A} for the alternative one.

We know from the general theory of \emph{evolution equations} (See for example \cite{Basarab-HorwathGuengoerLahno2013}),
the infinitesimal symmetry of the equation is generated by vector fields of the form
\be
X=\alpha(t)\partial_t+\beta(x,p,t)\partial_x +\gamma(x,p,t)
\partial_p +\eta(x,p,t)u\partial_u.
\el{VF}
We have excluded the trivial infinite-dimensional symmetry reflecting the linear superposition rule of the linear equations. The coefficients of the vector field are found from the requirement that the second prolongation $\mathsf {pr}^{(2)}X$ of $X$ annihilates the equation on its solution set. This gives us the following system of determining equations

\bes \lb{od} \begin{align}
% \nonumber to remove numbering (before each equation)
   &t^2\gamma_x-\beta_p=0,  \lb{od1}\\
\qquad &t(\beta_x-\gamma_p)-\alpha=0,  \lb{od2}\\
   &t^2 \gamma_{xx}-\gamma_{pp}-4t^2\gamma_t+2\eta_p=0,  \lb{od3}\\
   &t^2\beta_{xx}-\beta_{pp}-2t^2 \eta_x-4t^2\beta_t=0,  \lb{od4}\\
   &2\beta_x-\dot{\alpha}=0,  \lb{od5}\\
   &\eta_{pp}-t^2\eta_{xx}+4t^2\eta_t= 0. \lb{od6}
\end{align}
\ees

\sbs{Solving overdetermined system in \rf{od} }

In order to solve this overdetermined system we start from \rf{od5}  and find
$\beta=\frac{\alpha'(t)}{2}x+B(p,t)$. Eq.\rf{od2}  gives
$$\gamma=\frac{t\alpha'(t)-2\alpha(t)}{2t} p+C(x,t).$$
Substituting $\beta$ and $\gamma$ into the determining equations  \eqref{od}, we find $\alpha$, $B$ and $C$ satisfy
$$
\alpha(t)=a_2t^2+a_1t+a_0, \quad B=(c_1t+c_0)p+c_2 t+c_3, \quad C=\frac{c_1t+c_0}{t^2}x+C_0(t),$$
where $a_2$, $a_1$, $a_0$, $c_0$, $c_1$, $c_2$, $c_3$ are arbitrary integration constants. Finally $\eta$ has the form
$$\eta=-\frac{1}{2}\alpha''(t)x^2-2c_1xy-2c_2x+H(p,t),$$
where $H$ satisfies the system
$$tH_p=-4c_0x+2a_0tp+2t^3C_0'(t),  \quad H_{pp}+2t^2(2H_p+a_2)=0.
$$
From the compatibility of this system it follows $c_0=0$ and
$$
H=-\frac{4c_0xp}{t}+a_0p^2+2C_0'(t)t^2+\frac{a_0}{2t}-\frac{a_2t}{2}+c_4,
\quad C_0(t)=-\frac{c_5}{t}+c_6,
$$
where $c_4$, $c_5$ and $c_6$ are three further integration constants.

In conclusion, we find that $X$ depends on nine arbitrary integration constants which lead to  a basis of the symmetry algebra $\g$ as
\begin{equation}\label{Xi}
\begin{split}
% \nonumber to remove numbering (before each equation)
  &X_1=\gen t-\frac{p}{t}\gen p+(p^2+\frac{1}{2t})u\gen u, \\
  &X_2=2t\gen t+x\gen x-p\gen p, \\
  &X_3=t^2\gen t+xt\gen x-(x^2+\frac{t}{2})u\gen u,\\
	&X_4=tp \gen x+\frac{x}{t}\gen p-2xpu\gen u,\\
  &X_5=t\gen x-2xu\gen u,\\
  &X_6=-\frac{1}{t} \gen p+2pu\gen u,\\
  &X_7=\gen x,\quad X_8=\gen p,  \quad X_9=u\gen u.
\end{split}
\end{equation}
The Lie symmetry algebra $\g$ is identified as a 9-dimensional algebra with the structure
\begin{equation}\label{symalg}
\g=   \sl(2,\mathbb{R}) \oplus \so(1,1) \semiplus  \h_2\sim \langle X_1, X_2, X_3\rangle \oplus \langle X_4\rangle
\semiplus \langle X_5, X_6, X_7, X_8, X_9 \rangle,
\end{equation}
where $\h_2$ is the 5-dimensional Heisenberg algebra having the center $X_9$.
We see that the Lie symmetry algebra of \rf{psd} is isomorphic to that of the standard equation
\begin{equation}\label{standardpsd}
  u_t=u_{xx}-u_{yy}.
\end{equation}
This is seen from the commutation relations
\be
\begin{array}{llll}
     [X_1,X_2]=2X_1,  ~~&[X_1,X_3]=X_2,  ~~&[X_1, X_5]=X_7,   ~~&[X_1,X_8]=-X_6, \cr
     [X_2,X_3]=2X_3,  ~~&[X_2,X_5]=X_5,  ~~&[X_2,X_6]=-X_6,  ~~&[X_2,X_7]=-X_7, \cr
     [X_2,X_8]=X_8,   ~~&[X_3,X_6]=X_8,  ~~&[X_3,X_7]=-X_5, ~~&[X_4,X_5]=-X_8,\cr
     [X_4,X_6]=X_7,   ~~&[X_4,X_7]=X_6,   ~~&[X_4,X_8]=-X_5,   ~~&[X_5,X_7]=2X_9,  \cr
		                    &[X_6,X_8]=-2X_9.  &&
\end{array}
\el{comm}

\sbs{Isomorphy and point transformation}

The existence of such isomorphy indicates that Eq. \rf{psd} (after the change of scale $t\to t/4$) must be transformable to
its standard  form \rf{standardpsd} by a point transformation. This is indeed the case.
The transformation is given by
\be
t=t,  \quad x=x, \quad y=Y(p,t)=pt,  \quad Q(x,p,t)=t^{1/2}\exp[-\frac{tp^2}{4}] u(x,y,t).
\el{trans}
This is seen by the direct substitution
\be
L~ Q(x,p,t) = \sqrt{t} \,\exp[-\frac{tp^2}{4}]~
(\partial_t - \partial_{xx} + \partial_{yy})~ u(x,y,t)=0,
\el{Tr2Std}
so that Q defined by \rf{trans} is a solution of PSDE \rf{psd}, if $u$ is a solution of $u_t=u_{xx}-u_{yy}$.

Note that the above map of \rf{trans} is a special case of the map described below in \rf{gentr}.

\sbs{Generalizing the PSDE equation \rf{psd}}

Now we consider a more general equation
\be
u_t=a(t)u_{xx}+b(t)u_{yy},
\el{gen}
where $a(t)$ and $b(t)$ are positive nonzero arbitrary functions of $t$. By the analytic continuation $y\to iy$ it is equivalent to $u_t=a(t)u_{xx}-b(t)u_{yy}$.   We are interested in determining the case what should be the form of $b(t)$ for a given arbitrary $a(t)$ such that the equation can be mapped by a general coordinate transformation
\be
\tau=T(t),  \quad \xi=X(x,y,t), \quad \eta=Y(x,y,t), \quad u=U(x,y,t)v(\xi,\eta,\tau).
\el{gentr}
to the constant coefficient (standard) form $v_t=v_{\xi\xi}+u_{\eta\eta}$.

First of all, we can always set $a(t)=1$ by the transformation
$$\tau=\int a(t)dt, \quad u=v(x,y,\tau).$$ So we can start without loss of any generality with the equation
\be
u_t=u_{xx}+b(t)u_{yy}.
\el{cano}
The transformation
\be
\tau=t,  \quad \xi=x,  \quad \eta=\frac{y}{\sqrt{b(t)}}, \quad u=K(t)\exp[-\frac{b'}{b^2}\frac{y^2}{8}]v(x,\eta,t)
\el{sptr}
maps \rf{cano} to its standard form $v_t=v_{xx}+v_{\eta\eta}$ if and only if $K(t)$ and $b(t)$ satisfy
$$4b K'+Kb'=0,  \quad 2bb''=3b'^2.$$
Solving this system we find
$$b(t)=(b_1 t+b_0)^{-2},  \quad K(t)=k_0 b^{-1/4}=k_0 (b_1 t+b_0)^{1/2},$$ where $b_1$, $b_0$ and $k_0\ne 0$ are arbitrary integration constants. Hence we see that the factor $b(t)=t^{-2}$ obtained with the choice $b_1=1$, $b_0=0$ is the only function for which there is a transformation mapping the equation to the standard form.
The corresponding transformation (up to a nonzero multiplicative) is given by
$$u=t^{1/2}\exp[\frac{ty^2}{4}]v(x,ty,t).$$

This implies that for all equations with coefficients other than $b(t)=\text{const.}$ and $b(t)=t^{-2}$ the dimension of the symmetry algebra of the equation  will be less than nine.

We also performed  symmetry classification of \rf{gen} and found that the maximal symmetry algebra (9-dimensional) occurs if and only if either $b(t)$ is equal to a constant (the standard case) or $b(t)=t^{-2}$, otherwise the algebra is greatly reduced.
For any $b(t)$, the symmetry algebra is a five-dimensional Heisenberg algebra $\h_2$ and a basis is given by
\begin{equation}\label{h2x}
 \begin{split}
  &X_1 = t\gen x-\frac{xu}{2}\gen u,                  \quad X_2 =\gen x,     \\[2mm]
  &X_3 = (\int b(t)dt) \gen y-\frac{uy}{2}\gen u, \quad X_4 =\gen y,  \quad X_5 =u\gen u.
   \end{split}
\end{equation}
%\be
%\begin{array}{lll}
%% \nonumber to remove numbering (before each equation)
%  X_1 = t\gen x-\frac{xu}{2}\gen u,                  &\quad X_2 =\gen x, &    \\[2mm]
%  X_3 = (\int b(t)dt) \gen y-\frac{uy}{2}\gen u, &\quad X_4 =\gen y,  &\quad X_5 =u\gen u.
%\end{array}
%\el{h2}

The symmetry algebra is larger when $b(t)=b_0 t^{\alpha}$. Additional generator is
$$X_6=2t\gen t+x\gen x+(\alpha+1)y\gen y.$$
The symmetry algebra is further extended by three additional generators for the special choice $\alpha=0,-2$.

%%%%%%%%%%%%%%%%%%%%%%%%%%%%%%%%%%%%%%%%%%%%%%%%%%%%%%%%%%%%%%%%%%%%%%%%%%%%%%%%%%%%%%%%%%%%%%%%%%%%

\section{The 2-sided kernel of the PSDE \rf{psd}}  \label{kernel}

We now prove two simple propositions, which will help us determine the kernel of PSDE \rf{psd} in Sec. \ref{2sidedkernel}.

\sbs{Symmetry and factorization of solutions of the PSDE \rf{psd}}
\begin{prop}
\emph{Let $Q(x,p,t)$ is a solution of the PSDE \rf{psd},
then $Q(p,x, 1/t)$ is also a solution of \rf{psd}.}\label{prop1}
\end{prop}

\noindent \textbf{Proof.}  By assumption $Q$ satisfies $L(x,p,t)\, Q(x,p,t) = 0$. Clearly, this equation will still hold, if we change the names of the variables in $L$ and $Q$ simultaneously. Thus, we may write
\br 0 &=& L(x,p,t)\, Q(x,p,t) = L(p, x, 1/t)\, Q(p, x, 1/t) \crn
 &=&  \left[~\frac {\partial}{\partial (t^{-1})} -\frac 1 {4}\pd{^2}{p^2} +\frac 1 {4t^{-2}}\pd{^2}{x^2}~\right]Q(p, x, 1/t)\crn
 &=& -t^2 \left[~\frac {\partial}{\partial t}  + \frac 1 {4t^2}\pd{^2}{p^2} - \frac 1 {4}\pd{^2}{x^2}~\right]Q(p, x, 1/t) \crn
 &=& -t^2  L(x,p,t)\, Q(p, x, 1/t) \,.
\erl{prf}
Since the last expression in \rf{prf} is equal to zero, it follows that $Q(p, x, 1/t)$ is a solution of \rf{psd}. \qed   \\

\begin{cor} Let $f(x,t)$ be a solution of the 1-dimensional diffusion equation
\be
L_x f (x,t) := \left[~\dt-\frac 1 4 \pd{^2}{x^2}~\right]f(x,t) =0,
\el{x}
then $f(p,1/t)$ is a solution of the backward diffusion equation
\be
L_p f(p,1/t)\equiv \left[~\dt + \frac 1 {4t^2} \pd{^2}{p^2}~\right]f(p,1/t).
\el{p}
\end{cor}

\noindent \textbf{Proof.} Since $Q(x,p,t):=f(x,t)$ is a solution of \rf{x} it must also be a solution of \rf{psd}. Hence, its `dual' $Q(p,x,1/t)=: f(p,1/t)$
is also a solution of \rf{psd}. But since $f(p,1/t)$ is independent of $x$, it must be a solution of \rf{p}. \qed \\

Note that the above corollary can also be proved \emph{directly and easily} by using the same arguments as in the proof of proposition  \ref{prop1}.

\begin{prop}\label{prop2}
\emph{Let $f(x,t)$ and $g(p,t)$ be any two
solutions of the 1-dim equations \rf{x} and \rf{p}, respectively.
Then, their product}
\be
Q (x,p,t) := f(x,t) \, g(p, t),
\el{ufg}
\emph{is a solution of the pseudo-diffusion equation \rf{psd},}
\be
L~ [f(x,t) \, g(p,t)] \equiv
\left(\frac{\partial}{\partial t}-\frac{1}{4}\frac{\partial^2}{\partial x^2}
     +\frac{1}{4t^2}\pd{^2}{p^2}\right) [f(x,t) \, g(p,t)] =0.
\el{psdfg}
\end{prop}

\noindent \textbf{Proof.} Differentiating $fg$ by parts relative to $t$, we obtain
\br
\frac{\partial}{\partial t} [f(x,t)~ g(p,t)] &=&  \frac{\partial f(x,t)}{\partial t} \cdot g(p,t) +
f(x,t) \cdot \frac{\partial g(p,t)}{\partial t} \crn
&=&  \frac{1}{4}\frac{\partial^2 f(x,t)}{\partial x^2} \cdot g(p,t) -
f(x,t)\cdot \frac{1}{4t^2}\pd{^2 g(p,t)}{p^2 } \crn
&=& \frac{1}{4} \left[\frac{\partial^2}{\partial x^2}
     -\frac{1}{t^2}\pd{^2}{p^2}\right]~ [f(x,t)~ g(p,t)]\,. \qquad \qquad \qed \nn
\er

We can utilize the linearity of PSDE to construct solutions of \rf{psd} as sums
of the above product solutions \rf{ufg}
$$
Q (x,p,t)=\sum_i f_i(x,t)~g_i(p,t).
$$

\sbs{Obtaining the 2-sided kernel of the PSDE}  \lb{2sidedkernel}

Obtaining the kernel of the PSDE \rf{psd} requires careful analysis, since this equation
is diffusive in $x$, but infusive in $p$.
We shall need two different times $t_0 < t <t_1$ to define the initial conditions for the kernel.

It is well-known that the 1-dimensional heat equation has the following the \emph{Gaussian kernel}
\be
K_x \equiv K_0(x-x_0,t-t_0)=\frac{1}{\sqrt{\pi(t-t_0)}}\exp\Bigl[-\frac{(x-x_0)^2}{t-t_0}\Bigr],
\el{kx}
which is a solution of \rf{x}, defined by the initial condition
% (or fundamental solution)
\be
\lim_{\eps \to 0^+} K_0(x-x_0, \eps)=\lim_{\eps \to 0^+} \frac 1 {\sqrt{\pi\,\eps\,}}
\exp\Bigl[-\frac{(x-x_0)^2}{\eps}\Bigr] = \delta(x-x_0), \for \eps:=t-t_0 > 0.
\el{xdelta}
The corresponding solution of the \emph{`backward heat equation'} \rf{p} is given by
\br
K_p &\equiv& K_0(p-p_0,1/t-1/t_1)= \frac{1}{\sqrt{\pi(t^{-1}-t_1^{-1})}}
\exp\Bigl[-\frac{(p-p_0)^2}{t^{-1}-t_1^{-1}}\Bigr], \crn
&=& \sqrt{\frac{t t_1}{\pi(t-t_1)}} \exp\Bigl[+ \frac{t t_1 (p-p_0)^2}{t-t_1}\Bigr],
\erl{kp}
which becomes `anti-Gaussian' for $t> t_1$, since the exponential in \rf{kp} will blow
up in the limit $t \to t_1^+$ for $p\ne p_0$. Therefore, in order to impose the usual delta function  $\delta(p-p_0)$ on the initial condition for the $K_p$-kernel, we must choose $t_1> t$, \ie  $t$ should approach $t_1$ from below.

\comm{ But it is inconsistent to demand that
$t>t_0$ for $K_x$ \emph{and} $t <t_0$ for $K_p$. So, we must another time, call it $t_1 > t > t_0$ and let $t$ approach $t_1$ from below.
Some people may like to use one parameter $\eps > 0$ to define the two bounds, $ t_0 =t-\eps < t < t_1=t+\eps$, and impose the limit $\eps \to 0$.
endcomm }
Thus, we define the p-kernel as a solution of \rf{p} which satisfies the initial condition
\be
\lim_{t \to t_1^-} K_p(p-p_0, 1/t-1/t_1) = \delta(p-p_0).
\el{limkp}
Finally, we obtain the kernel of the PSDE \rf{psd}, according to proposition 2, as a product of the above two kernels $K_x$ and $K_p$:
\br
K(x-x_0,t-t_0; p-p_0,t_1-t)&=&\frac{tt_1}{\pi\sqrt{(t-t_0)(t_1-t)}}\exp\Bigl[-\frac{(x-x_0)^2}{t-t_0}- \frac{t t_1 (p-p_0)^2}{t_1-t}\Bigr],\crn
&& \for t_0  < t < t_1.
\erl{kxp}
The above algorithm for obtaining the kernel in \rf{kxp} was already described by Daboul in 1995 \cite{dab2} and then in 1996 \cite[Eq.(31)]{daboul}.
He called the kernel in \rf{kxp} the \emph{`forward-backward  propagator'}. This kernel may also be called  a \emph{`2-sided kernel'}, since its initial conditions are defined by $t$ approaching $t_1$ and $t_0$ from below and from above, $t_0 < t < t_1$.

\section{List, table and properties of the symmetry operators $\g=<A_i>$}

% which are derived in Appendix A by the alternative approach}

In Eq.\rf{ai} below we list the nine infinitesimal symmetry  operators $\{A_i\}$,  which are related to the $\{X_i\}$ in \rf{Xi} (see Remarks \ref{remark1} and \ref{remark2}).
We can obtain the $\{A_i\}$ from the $\{X_i\}$ by replacing every $u\gen u$ in \rf{Xi} by $-1$ in \rf{ai}.

We derive the  $\{A_i\}$ in details by a simplified procedure in Appendix \ref{A}. We take a different approach and show explicitly that
$A_i$ of the form given in Eq.\rf{a} in the sense that the $A$'s commute with the differential operator $L$ on the solution set.
From this invariance requirement  all the coefficients
$\alpha, \beta, \ga, \eta$ in \rf{a} can be determined as explicit functions of
9 parameters $c_1,c_2,\cdots,c_9$, which are essentially arbitrary
integration constants.
Thus, we can write $A$ as a sum of the form
$$
A=\sum_{i=1}^9 c_i A_i,
$$
where $A_i$ are the following infinitesimal generators:
\be
\begin{array}{ll}
A_1= \partial_t - \frac{p}{t}\,p\partial_p - p^2 -\frac{1}{2t}  =:K_-, & \cr
A_2=2t\partial_t+x\partial_x-p\partial_p =: K_0, &\quad\\
A_3=t^2\partial_t + tx\partial_x + x^2+\half t = K_+, & \\[3mm]
A_4=pt\partial_x+\frac{x}{t}\partial_p+2xp=:J, & \text{pseudo-rotation} \\[3mm]
A_5=2x + t\partial_x\equiv R(2x, t), &\quad \\
A_6=2p+ \frac{1}{t}\partial_p\equiv R(2p, 1/t) \simeq -X_6, & \\
A_7=\partial_x, & \cr
A_8=\partial_p, & \\
A_9=1 \simeq -X_9. &
\end{array}
\el{ai}
The commutation relations among the various $A_i$ are summarized in Table 1.
$$
\ba{|r|ccc|c|cccc|c|}
\hline
        &  K_0  & K_- & K_+ & J & D_1^-& D_1^+ & D_2^+& D_2^- & L \\
				&  A_2  & A_1 & A_3 & A_4 & A_7  & A_5   & A_8  & A_6   & L \\ \hhline
%---------------------------------------------------------------------------				
K_0 = A_2 & 0   &-2K_-& 2K_+ & 0  & -A_7 & A_5   & A_8  & -A_6    & -2 L \\ \hline
K_- = A_1 &    &   0 &  K_0 & 0   & 0    & A_7   & A_6  & 0      & 0    \\ \hline
K_+ = A_3 &    &     &  0   & 0   &-A_5  &  0    &  0   &-A_8    &-2tL  \\ \hline
J = A_4 &    &     &      & 0   & -A_6 & -A_8  & -A_5 & -A_7   & 0   \\ \hline
D_1^-=A_7 &   &     &      &      & 0    & 2    & 0    & 0      & 0  \\ \hline
D_1^+=A_5 &   &     &      &      &      &  0    & 0    & 0      & 0  \\ \hline
D_2^+=A_8 &   &     &      &      &      &       & 0    & 2     & 0  \\ \hline
D_2^-=A_6 &   &     &      &      &      &       &      & 0      & 0    \\
\hline
\ea
$$
\begin{center}
Table 2: The commutation relations among the infinitesimal generators
$A_i$ in \rf{ai}.
$<K_0, K_\pm> \simeq \su(1,1) \simeq \sl(2,\mathbb{R})$ commute with $J \simeq \so(1,1)$.
\end{center}

\sbs{Symmetries of the $A_i$}

It is interesting to note the symmetry of the PSD operator $L$ in \rf{psd} and the consequent
symmetries among the $A_i$ under the exchange of the variables
$x  \Leftrightarrow p$ and $t  \Leftrightarrow 1/t $:
\be \begin{array}{lll}
&~ L(x,p,t) = - L(p,x,1/t) & \cr
   &A_3(x,p,t) = - A_1(p,x,1/t),  \qquad   &A_2(x,p,t) = - A_2(p,x,1/t), \comm{as it should !!} \cr
	&A_4(x,p,t) = A_4(x,p,1/t),  \qquad    &  \cr
	&A_5(x,p,t) = A_6(p,x,1/t),  \qquad   &A_7(x,p,t) = A_8(p,x,1/t).
	\ea
	\el{sym}
The subalgebra of operators $<A_1, A_2, A_3>$ in \rf{ai} is isomorphic to the Lie algebra $\sl(2,\mathbb{R})$, so we have
identified them in \rf{ai} as such. They commute with $ J \equiv A_4$.

Each triplet $(A_6, A_8, A_9)$ and $(A_5, A_7, A_9)$ commutes like a 3-dimensional Heisenberg algebra $\h_1$,
and together ($A_5,\ldots,A_9$) they span a 5-dimensional Heisenberg algebra $\h_2$.

Moreover, we have
\be
[A_i, A_j] = A_k \in \h_2, \qquad i=1,\ldots,9, \quad j,k=5,\ldots,9,
\el{aiaj}
so that $\h_2$ is an ideal of the whole algebra $\g$.
This identification will be further discussed  in Appendix \ref{B}.

\begin{rmk}\label{remark1}
Note that the operator $A$ and the vector field $X$ are related by the following relation:
the $u$-coefficient of $X$ is equal to $-u$ times the last function in $A$. In other words, the $X_i$ yields the corresponding $A_i$ if we replace every $-u\gen{u}$ in the $X_i$ by 1. Though the two approaches gave rise to
the same number of generators in this case, because of the restricted form of $A$, the method of linear operators
may not give all the symmetries in general.
\end{rmk}

\begin{rmk}\label{remark2}
If we set $\eta=-\phi(x,p,t)$ in \rf{a} then we have $W\equiv -Au=[\phi u-(\alpha u_t+\beta u_x+\gamma u_p)]$ and
$\hat{X}= W \, \gen u$ will be a symmetry vector field in evolutionary form with the characteristic function $W$ \cite{olver}.
$\hat{X}$ is equivalent to the usual vector field
$$
X=\alpha\gen t+\beta\gen x+\gamma\gen p+\phi u\gen u.
$$
\end{rmk}

%%%%%%%%%%%%%%%%%%%%%%%%%%%%%%%%%%%%%%%%%%%%%%%%%%%%%%%%%%%%%%%%

\section{The groups generated by the infinitesimal symmetry
operators $A_i$}

\sbs{How to compute $\exp [\la A(x,p,t)] f(x,p,t)$ ?} \label{groups}

To compute $\exp [\la A(x,p,t)] f(x,p,t)$ we start we the following Ansatz:
\be
e^{\la A}  f(x,p,t)=\sigma(x,p,t,\la)f(X,P,T),
\el{action}
where the functions $X$, $P$, $T$ (which in general depend on $x,p,t,\la $) and $\sigma$ are uniquely determined from the solution of the initial value problem for the system of ordinary differential equations
\bes \label{ODEs} \begin{align}
& \frac{dX}{d\la}=\alpha(X,P,T),  \qquad \frac{dP}{d\la}=\beta(X,P,T),  \qquad \frac{dT}{d\la }=\gamma(X,P,T), \\[2mm]
& \frac{d\sigma}{d\la }=\sigma(x,p,t,\la)\,\eta(X,P,T),   \\[2mm]
& X(x,p,t,0)=x,  \quad P(x,p,t,0)=p,  \quad T(x,p,t,0)=t, \quad \sigma(x,p,t,0)=1.
\end{align}
\ees

For example, the action of the pseudo-rotation $\exp[\la A_4]$ in \rf{gi} on the solution $Q(x,p,t)$ is obtained by solving the system
\bes \label{G4} \begin{align}
   &\frac{dX}{d\la }=T P, \qquad \frac{dP}{d\la }=\frac{X}{T},\qquad \frac{dT}{d\la }=0, \\[2mm]
   &\frac{d\sigma}{d\la }=2\sigma XP, \lb{G4b} \\[2mm]
   &X(x,p,t,0)=x,  \quad P(x,p,t,0)=p,  \quad T(x,p,t,0)=t, \quad \sigma(x,p,t,0)=1.
\end{align}
\ees

It follows that    $T=t$ so that the first two linear ODEs can be solved easily in the form
$$X=x \cosh \la +tp\sinh \la , \quad P=p \cosh \la +\frac{x}{t}\sinh \la .$$ We substitute
$(X,P)$ in \rf{G4b} and find the initial value problem
$$\frac{d\sigma}{d\la }=\sigma(2xp\cosh 2 \la +\frac{x^2+t^2p^2}{t}\sinh \la ),\quad \sigma(x,p,t,0)=1.$$
This problem is easily solved in the form
$$\sigma(x,p,t,\la )=\exp\left[\sinh^2 \la \left(\frac{x^2}{t}+ ~p^2
t\right)+x p ~\sinh(2\la )\right].$$
Finally we use the formula \rf{action} to find the transformation of the solution under the pseudo-rotation operator
$$
Q'=e^{\la A_4}Q(x,p,t)=\exp\left[\sinh^2 \la \left(\frac{x^2}{t}+ ~p^2
t\right)+x p ~\sinh(2\la )\right]Q(X,P,t).
$$
This relation implies that $Q'$ solves the equation under study whenever $Q(x,p,t)$ does.

In the following we list all the
one-parameter group actions $G_i\equiv \exp[\la A_i]$ generated by $A_i$
on any solution $Q(x,p,t)$ of \rf{psd}:
\be
\begin{array}{rll}
G_1~Q(x,p,t) \comm{&=Q\left(x,\frac {p \,t} {t+\la }, {t+\la } \right)
~\frac 1 {\sqrt{1+\la /t}}\exp\left[-\frac{\la t p^2}{t+\la }\right]
&\mbox{(conformal in $p$ and $t$)}\\[2mm]  }
&=Q\left(x,\frac {p \,t} {t+\la }, {t+\la } \right)\frac 1 {\sqrt{1+\la /t}}
\exp\left[-\frac{p^2}{1/\la + 1/t}\right] &\mbox{(conformal in $p$ and $t$)}\\[2mm]
G_2~Q(x,p,t)&=Q(e^\la x, e^{-\la }p,e^{2\la }t) &\mbox{(scale symmetry)}\\[2mm]
G_3~Q(x,p,t) \comm{&= Q\left(\frac x {1+\la t}, p, \frac t {1+\la t}\right)
~\frac 1 {\sqrt{1+\la t}}\exp\left[-\frac{\la x^2}{1+\la t}\right] &
\mbox{(conformal in $x$ and $t$)}\\[3mm] }
&= Q\left(\frac x {1+\la t}, p, \frac t {1+\la t}\right)
~\frac 1 {\sqrt{1+\la t}}\exp\left[-\frac{x^2}{1/\la + t}\right] &
\mbox{(conformal in $x$ and $t$)}\\[3mm]
%A_4=pt\partial_x+\frac{x}{t}\partial_p+2xp=:J^0_T,&\\[2mm]
G_4~Q(x,p,t) &= Q(x~\cosh \la + t p ~ \sinh \la , ~ p ~ \cosh \la +\frac x t ~ \sinh \la , t) &\mbox{(`pseudo-rotation')} \\[2mm] %\crn
& \x \exp\left[\sinh^2 \la \left(\frac{x^2}{t}+ ~p^2 t\right)+x p ~\sinh(2\la )\right]  & \\[2mm]
G_5~Q(x,p,t)&= Q(x+\la t,p,t) ~\exp[2\la x + \la ^2 t] & \mbox{(Galilean boost in $x$)}\\[2mm]
G_6~Q(x,p,t)&= Q(x,p+\la / t,t) ~\exp[2\la p + \la ^2 /t]&
\mbox{(Galilean boost in $p$)}\\[2mm]
G_7~Q(x,p,t)&=Q(x+\la ,p,t) &\mbox{($x$-translation)}\\[2mm]
G_8~Q(x,p,t)&=Q(x,p+\la ,t) &\mbox{($p$-translation)}\\[2mm]
G_9~Q(x,p,t)&=Q(x,p,t)~e^{\la }. &\mbox{(phase factor)}
%\end{align}%
\end{array}
\el{gi}

\section{Applications of the symmetry generators $A_i$ and their Exponentiation $G_i$}

By definition, symmetry operators $A_i$ map solutions to other solutions $u \mapsto A_i u$. Consequently, their exponentiation $G_i=\exp[\ga A_i]$ also map solutions to other solutions $u \mapsto G_i u$.

\sbs{Generating heat polynomials by repeated application of $A_5$}

Applying powers of $A_5$ on the trivial solution $u=1$ yields %immediately the \emph{heat polynomials}, as follows
\be
A_5^n \cdot 1 = (2x + t \,d/dx)^n   \cdot 1 = v_n(2x,t)\,,
\el{a5n}
where $v_n(x,t)$ is called the \emph{heat polynomial of degree $n$} \cite{widder}\,:
\be
v_n(x,t) := \sum_{k=0}^{\floor{n/2}} \frac {n!}{(n-2k)! \, k!}~ x^{n-2k}\, t^k.
\el{vn}
We obtain $v_n(2x,t)$ in \rf{a5n} instead of $v_n(x,t)$, because the PSDE \rf{psd} has the factor $1/4$. In particular,
\be
v_0=1, \quad v_1(2x,t)= 2 x, \quad  v_2(2x,t)= 4 x^2 + 2t , \quad v_3(2x,t)= 8 x^3 + 12 x t .
\el{v0}

For $t=-1$ our $v_n(2x,t)$ yield the
\emph{Hermite polynomials} $H_n(x)$ used by physicists. Thus,
\be
H_n(x) = v_n(2x,-1) = (2x - d/dx)^n   \cdot 1 =  (-1)^n e^{x^2} \frac {d^n}{dx^n} e^{-x^2},
\el{hn}
so that
\be
H_0(x) = 1, \quad H_1(x) = 2 x, \quad  H_2(x) = 4 x^2 - 2 , \quad H_3(x)= 8 x^3 - 12 x .	
%H_4(x)	=	16x^4-48x^2+12.
\el{h0}
Actually, the heat and Hermite polynomials follow as special cases of the following polynomial
$\tilde{H}_n(\al,\beta; x)$ which is obtained by applying the operator $R(\al, \beta; x)$ on the unity 1:
\be
R(\al, \beta; x)^n \cdot 1 := (\al x -\beta \frac {d}{dx})^n \cdot 1 =: \tilde{H}_n(\al,\beta; x).
\el{ghp}
In particular,
\br
\widetilde{H}_ 0 (x) &=& 1, \quad \widetilde{H}_ 1 (x) = \al x, \quad \widetilde{H}_ 2 (x) =
  \al^2 x^2 - 2 \beta, \\
	\widetilde{H}_ 3 (x) &=& \al^3 x^3 - 6 \al \beta x\, \quad	\widetilde{H}_ 4 (x) =  \al^4 x^4 - 12 \al^2 \beta x^2 + 12 \beta^2, \\
 \widetilde{H}_ 5 (x) &=& \al^5 x^5 - 20 \al^3 \beta x^3 + 60 \al \beta^2 x.
\erl{ghp15}
The operator $R(\al, \beta; x)$ was introduced by Daboul and Mizrahi to prove a new sum rule
\cite[Eqs. 4 \& 5]{dmh} which was encountered in squeezing formalism. The
$\widetilde{H}_n(\al,\beta; x)$ were called \emph{generalized Hermite polynomials} ($\cal{GHP}$)
These polynomials yield different polynomials of degree $n$ as special cases:
\be
\ba{lrlrll}
\cal{GHP} & \widetilde{H}_n(\al, \beta; x) &:=& R(\al, \beta; x)^n \cdot 1 &= (\al x - \beta\frac {d}{dx})^n \cdot 1,& \\[2mm]
\text{Hermite polynomials}& H_n(x,t) &:=& R(2, 1; x)^n \cdot 1 &= (2x - \frac {d}{dx})^n \cdot 1= (-1)^n e^{x^2}
\frac {d^n}{dx^n} e^{-x^2},& \\[2mm]
\text{Heat polynomials}& v_n(x,t) ~&:=& R(1, -t; x)^n \cdot 1 &= (x + t \frac {d}{dx})^n \cdot 1,& \\[2mm]
\text{Simple powers} &(\al x)^n &:=& R(\al, 0; x)^n \cdot 1, &  &\\[2mm]
\text{Raising operators} &a^\dg &:=& R(\frac 1 {\sqrt{2}x_0}, \frac{x_0}{\sqrt{2}}; x)
&= \frac 1 {\sqrt{2}}(x/x_0 - x_0 \frac {d}{dx}), &   % \\[2mm]
\ea
\el{}
where $ x_0 := \sqrt{\hbar/(m\om)}$ and $a^\dagger$ is exactly the same as the raising operator of the standard harmonic oscillator in \rf{aa}.

\sbs{Examples of solutions obtained by the group symmetry operators}

The symmetry group operators $G_i$ in \rf{gi} yield new solutions from
known ones. Here are some examples:
\bigskip

\sbb{Generators of the $\cal{GHP}$ and heat polynomials}

In the previous Subsection we applied powers of $A_5$ to unity and obtained heat polynomials.
We now show that the exponential of $A_5$ yields its generating function.

According to \rf{gi}, by applying the \emph{Galilean transformation} $G_5 =\exp[\ga A_5 (x,t)]$
on the trivial solution $Q=1$, we obtain
\be
G_5(\ga )\cdot 1 = \exp[2\ga x+\ga ^2t] .
\el{expsol}
We can now \emph{easily} prove that the above exponential solution in \rf{expsol} is in fact the generator of our
heat polynomial solutions $v_n(2x,t)$:
\br
\exp[2\ga x+\ga ^2 t] &=& G_5(\ga )\cdot 1 = \exp[\ga A_5] \cdot 1 \crn
&=& \sum_{n=0}^\infty \frac {\ga ^n} {n!}\, A_5^n \cdot 1 = \sum_{n=0}^\infty \frac {\ga ^n}{n!}\, v_n(2x,t).
\erl{heatpol2}
The equality in \rf{heatpol2} is derived in \cite[Eq.(1.8)]{widder} in a more complicated way, by using convolution of $x^k$ with the kernel $k(x,t)$.

Similarly, by setting $t=-1$ in \rf{heatpol2}, we obtain immediately the generating function
\cite[Eq.(1.10)]{widder} of the Hermite polynomials defined in \rf{hn}:
\be
\exp[2\ga x -\ga ^2] = \sum_{n=0}^\infty \frac {\ga ^n}{n!}\, H_n(x).
\el{expheat}
One can check that the generating function of the
Generalized Hermite Polynomials $\cal(GHP)$ is given by
\be
\exp[\ga \al x -\beta \ga ^2] = \sum_{n=0}^\infty \frac {\ga ^n}{n!}\, (\al x -\beta\, d/dx)^n \cdot 1
= \sum_{n=0}^\infty \frac {\ga ^n}{n!}\, \widetilde H_n(\al,\beta; x) .
\el{ghp-generate}

\sbb{Deriving the kernels by the conformal maps $G_1$ and $G_3$}

Mathematicians and theoretical physicists often ignore dimensions and units, for the sake simplicity.
If we allow in the present paper a single dimension, the time $[t]$, the if follows from the PSDE \rf{psd}
that $x^2$ and $p^2$ have the dimensions $[t]$ and $[1/t]$, respectively.
This enables us to understand that the infinitesimal generators $A_1$ and $A_3$ have the dimensions
$[1/t]$ and $[t]$, respectively, which force their expansion parameters
$\ga _1$ and $\ga _3$ to have the inverse dimensions. Thus, by noting \rf{gi} and equating $\ga _1= -t_1$ in $G_1$, but $\ga _3=-1/t_0$ in $G_3$, we immediately obtain the $p$-and $x$-kernels, as follows:
\br
G_1(\ga_1)\cdot 1 &=& \exp[ \ga _1 A_1(x,t)] \cdot 1
= \frac 1 {\sqrt{1 +\ga _1/ t}}~\exp\left[-\frac{p^2}{1/t + 1/\ga _1}\right] \\[2mm]
&=& \frac t {\sqrt{t-t_1}}~\exp\left[-\frac{t_1 t p^2}{t_1 - t}\right]
\erl{kg1}
and
\br
 G_3(\ga _3)\cdot 1 &=& \exp[ \ga _3 A_3(x,t)] \cdot 1
= \frac 1 {\sqrt{1+\ga_3 t}}~\exp\left[-\frac{x^2}{t+ 1/\ga_3 }\right] \\[2mm]
&=& \frac 1 {\sqrt{1- t/t_0}}~\exp\left[-\frac{x^2}{t -t_0}\right].
\erl{kg3}
We see that except for an imaginary constant factor \emph{``i``} the conformal map $G_1(t_1)$ and
$G_3(-1/t_0)$ yield the two kernel immediately when applied to one.

\sbb{Obtaining the squeezed thermal distribution by again applying conformal mapping}

Hence, according to proposition 2, the product of the above two solutions
\br
Q_{th}(x,p,t) &=& G_3(x,t,\ga )\cdot 1 \x G_1(p,t;\ga )\cdot 1 \crn
&=&\frac 1 {\sqrt{(1/\ga + t)(1/\ga +1/t)}}
\exp\left[-\frac{x^2}{1/\ga + t}-\frac{p^2}{1/\ga +1/t}\right]
\erl{thsqueezed}
must be a solution of the PSDE \rf{psd}.

The solution in Eq. \rf{thsqueezed} yields the \emph{\textbf{squeezed} thermal distribution} given in
\cite[Eq.(82)]{dmm}, if we choose $1/\ga =2\bar{n}+1$, where $\bar{n}$ is the average
number of photons:
\be
Q_{th}(x,p,t) = \frac 2 {\sqrt{(2\bar{n}+1 + t)(2\bar{n}+1+1/t)}}
\exp\left[-\frac{x^2}{2\bar{n}+1 + t}-\frac{p^2}{2\bar{n}+1 +1/t}\right].
\el{squeezed-therm}
For $t=1$ (no squeezing) the above $Q_{th}(x,p,t)$ reduces to the well-known un-squeezed thermal
distribution \cite[Eq.(82)]{dmm}, which is symmetric in $x$ and $p$\,:
\be
Q_{th}(x,p,1) = \frac 1 {\bar{n}+1}~\exp\left[-\frac{(x^2 + p^2)}{2\bar{n}+2}\right] \for t=1.
\el{therm}

It is interesting to note that the integral
$$
\int_{-\infty}^\infty\frac 1 {\sqrt{1+\ga t}}~\exp\left[-\frac{\ga x^2}{1+\ga t} \right] dx =
\int_{-\infty}^\infty\frac 1 {\sqrt{1+\ga /t}}~\exp\left[-\frac{\ga p^2}{1+\ga /t} \right] dp = \sqrt{\pi}
$$
is independent of $t$. This fact yields a direct proof that \emph{squeezing keeps the integral over
the 2-dimensional distribution function $Q(x,p,t)$ invariant.}
\bigskip

\sbb{Applying different generators in succession}
We can obtain more complicated and perhaps more interesting solutions, if we apply different
symmetry operators one after another.

For example, by applying $G_3(\ga )$ on the heat polynomial solution $v_n(2x,t)$ of the PSDE, we obtain:
\be
G_3(\ga )\, v(2x,t)= v_n \left(\frac {2x} {1+\ga t},\frac t {1+\ga t}\right)
~\frac 1 {\sqrt{1+\ga t}}\exp\left[-\frac{\ga x^2}{1+\ga t}\right]\,.
\el{g3h}
This solution yield a \emph{real} distribution whose integral over the $x$-axis is finite, if $ (1+\ga t) > 0$ and $ \ga > 0$.
Thus, the transformation $G_3(\ga )$ enables us to obtain the normalizable solution \rf{g3h} from
the non-normalizable heat polynomial, $v_n(2x,t)$. This is possible since the map $G_3$ is not unitary.

\begin{appendices}
\section{Derivation of the symmetry generators $A_i$}  \label{A}

In this Appendix we look for the most general operator of the form
$ A =\alpha~ \partial_t+\beta~\partial_x +\gamma~\partial_p +\eta$, as defined
in \rf{a}, which satisfies the condition $[L, A] -\xi(x,p,t) L =0$ in \rf{rq}. Substituting
\rf{a} into \rf{rq}, we obtain
\be
[L, A] - \xi L= \left\{ ~[L,\al]\pdt+ \al [L, \pdt]~\right\} +
[L, \beta ]\pdx+ [L, \ga ]\pdp+ [L, \eta ] -\xi L =0.
\el{la}
In order to calculate the commutator of $L$ with any function $f(x,p,t)$, we proceed
as follows:
\comm{\be \left[\pd{^2}{x^2}, f(x,p,t)\right]=
\dx[\dx,f]+ [\dx,f]\dx=
\dx \pd{f}{x}+\pd{f}{x}\dx=
f_{xx}+2f_x \pdx.  \ee \endcomm}
\be
\left[\pd{^2}{x^2}, f(x,p,t)\right] \psi= (f\psi)_{xx} - f \psi_{xx} = f_{xx} \psi + 2 f_x \psi_x = [f_{xx}  + 2 f_x \px] \psi.
\ee
Similarly, we obtain
\be
\left[\pd {^2}{p^2}, f(x,p,t)\right]=
f_{pp}+2f_p \pdp, \quad
\left[\dt, f(x,p,t)\right]=f_t.
\ee
Using the above relations we obtain
\be
[L, f(x,p,t)]=\left[\dt - \frac 1 4 \pd{^2}{x^2}+ \frac 1 {4t^2} \pd{^2}{p^2}, f(x,p,t)\right]=
L(f) - \half f_x \pdx +\frac 1{2t^2} f_p \pdp,
\el{lf}
where we use $L(f)$ to denote the following {\em function}
\be
L(f) \equiv f_t-\frac 1 4 \left(f_{xx}-\frac 1 {t^2}f_{pp}\right).
\ee
In contrast, we use $Lf$ to denote the {\em operator}
$$ L f \psi :=([L,f] + fL)\psi.$$
Substituting the expression \rf{lf} into \rf{la}, we obtain
\br
[L, A ]- \xi L &=&
\left\{\left( L(\al) -\half\al_x \pdx +
\frac 1{2 t^2} \al_p \pdp \right)\pdt +\frac {\al}{2t^3} \pdp^2 \right\} +\cr
&\,& \left(L(\beta)
-\half\beta_x \pdx+\frac 1{2 t^2} \beta_p \pdp
\right)\pdx + \left(L(\ga)-\half\ga_x \pdx+\frac 1{2 t^2} \ga_p
\pdp \right)\pdp  \cr
&\,& +
\left(L(\eta) -\half\eta_x \pdx+\frac 1{2 t^2} \eta_p \pdp
\right)-\xi L \nn \\
&=& L(\eta)
+(L(\beta )-\half \eta_x)\pdx +
(L(\ga) +\frac 1 {2t^2} \eta_p)\pdp +
(L(\al) -\xi)\pdt \nn \\
&\,& -\half \al_x \pdx\pdt+\frac
1{2t^2}\al_p\pdp\pdt
+\left(\frac1{2t^2}\beta_p-\half \ga_x\right)\pdx\pdp
\nn\\
&\,&
+\left(-\half \beta_x+\frac \xi 4\right)\pdx^2
+\frac 1 {2t^2}\left(\ga_p+\frac \al t -\half \xi
\right)\pdp^2=0.
\er
By setting the coefficients of the
partial derivatives equal to zero, we obtain the
following conditions
\bes \lb{cond}
\begin{align}
L(\eta) &= 0, \quad L(\beta) =\half \eta_x, \quad L(\ga)=-\frac 1 {2t^2} \eta_p, \quad
L(\al) =\xi,  \lb{cond1}  \\
\al_x&= \al_p=0, \lb{cond2} \\
\beta_x &= \half \xi, \quad \ga_p=\half \xi - \frac \al t \,,\lb{cond3} \\
\beta_p &= t^2 \ga_x. \lb{cond4}
\end{align} \ees
From \rf{cond2} we conclude that $\al$ is independent
of $x$ and $p$, so that $\al=\al(t)$. This conclusion can be
proved in general for a \emph{general evolution equation} \cite{Basarab-HorwathGuengoerLahno2013}, so that it is
already written as $\al(t)$ in the ansatz for the vector field in \rf{VF}.

Hence, the condition $L(\al)=\xi$ in \rf{cond1} yields
\be   \xi = L(\al)=\al_t\equiv \dot \al,
\el{alt}
so that $\xi=\xi(t)$ is a function of $t$
only. In turn, \rf{cond3} yields similarly $\beta_x (t)$ and $\ga_p(t)$. Hence,
\be
\beta_{xx}= \beta_{xp}= \gamma_{px}= \gamma_{pp}=0.
\el{bb}
By noting \rf{bb} the condition \rf{cond4} yields
\be
\beta_{pp}=\gamma_{xx}=0.
\el{bb2}
The conditions  \rf{bb}
and \rf{bb2} tell us that  $\beta$ and $\gamma$ can have at most \emph{linear}
terms in $x$ and $p$, with time-dependent coefficients. Moreover, by noting $\beta_p= t^2 \ga_x$ from
\rf{cond4}, we can write the following ansatz
\bes \lb{bg}
\begin{align}
\beta &= \half \xi(t)  x + \sigma (t) p + \ga (t),   \lb{bg1} \\
\ga &= \left(\half \xi -\frac \al t \right) p + \frac {\sigma(t)}{t^2} x+\rho(t). \lb{bg2}
\end{align}
\ees
So far $\sigma,\ga $ and $\rho$ are arbitrary
functions of $t$, which will be determined shortly.

By noting \rf{bb} and \rf{bb2} we conclude from \rf{cond1} that
\be
\beta_t= L(\beta)=\half \eta_x
, \qquad
\gamma_t=L(\gamma)=-\frac 1 {2t^2} \eta_p,
\el{et1}
which in turn, by noting \rf{bg}, we can conclude that
\be
\eta_{xxx}=\eta_{xxp}=\eta_{xpp}=\eta_{ppp}=0.
\el{ett}
The above conditions in \rf{ett} tell us that $\eta$ {\em is a  polynomial
solution of degree 2 in $x$ and $p$}. Moreover, the condition $L(\eta)=0$ in \rf{cond1} tells us that $\eta$
is a solution of PDE \rf{psd}. Therefore,  we can expand $\eta$
as a sum of heat-polynomial solutions of \rf{psd} as follows:
\br
\eta &=&\frac 1 4 [c_3 v_2(2x,t)- c_1 v_2(2p,1/t)] +
\frac {c_4} {2} v_1(2x) v_1(2p)+c_5 v_1(2x)+c_6 v_1(2p)+c_9 \cr
 &=& c_3 (x^2 +\half t)-c_1 (p^2+ \frac{1}{2t})
+ 2 c_4 xp+2 c_5 x+ 2 c_6 p + c_9,
\erl{eta}
where we used $v_1(2x)=2x$ and $ v_2(2x,t)=4x^2+2t$.

From \rf{et1} and \rf{eta}, we obtain
\bes  \lb{bega}
\begin{align}
\beta_t &=\half \xi_t x+\s_t p+\ga _t= \half\eta_x \cr
&= c_3x+c_4p+c_5, \lb{bega1}\\[2mm]
\ga_t &= \left(\half \xi -\frac \al t \right)_t p +
\left(\frac {\sigma(t)}{t^2}\right)_t ~x+\rho_t= -\frac 1 {2t^2} \eta_p \cr
&= \frac 1 {t^2}\left(c_1p-c_4x-c_6\right). \lb{bega2}
\end{align}
\ees
Eq.\rf{bega1} yields
\bes  \lb{cs1} \begin{align}
\sigma_t=c_4 &\quad \quad \Rightarrow \quad \s=c_4 t,     \lb{cs1a}\\[2mm]
\ga _t=c_5    &\quad \quad \Rightarrow \quad  \ga =c_5 t+c_7, \lb{cs1b}\\[2mm]
\xi_t=2c_3   &\quad \quad \Rightarrow \quad  \xi=2c_3t+2c_2. \lb{cs1c}
\end{align} \ees
Eq.\rf{bega2} yields two new conditions
\be
\ba{rll}
(\half \xi -\frac  \alpha t)_t &= c_1t^{-2} &\quad \Rightarrow \quad
\alpha=c_3t^2+2c_2t+  c_1, \\[2mm]
\rho_t&=-c_6t^{-2}&  \quad \Rightarrow \quad \rho=c_6t^{-1} + c_8,
\ea
\el{cs2}
where we substituted the expression for $\xi$ in \rf{cs1c}.
Substituting the above expressions into \rf{bg} we obtain the final expressions for
$\beta$ and $\gamma$:
\br
\beta &=& (c_3t +c_2)x +c_4 t p+c_5t+c_7,    \lb{bet}\\[2mm]
\gamma &=& -(c_2+\frac{c_1}{t})\,p+c_4\frac{x}{t}+\frac{c_6}{t}+c_8.
\erl{ga}
Substituting into the general ansatz for $A$ in \rf{a} the expressions $\alpha, \beta, \ga$ and $\eta$ (which we determined in Eqs. \rf{cond2},\rf{bet},\rf{ga}, and \rf{eta}, respectively),
we obtain
$$
 A =\alpha(t)  \partial_t+\beta\partial_x +\gamma\partial_p +\eta =\sum_{i=1}^9 c_i A_i\,.
$$
We see that $A$ depend on 9 independent constants $\{c_i\}$, whose coefficients $A_i$ yield the 9 symmetry generators, listed
explicitly in \rf{ai}.

\section{Algebraic structures}\label{B}

%\urb{https://en.wikipedia.org/wiki/Solvable_Lie_algebra}
%\urb{https://en.wikipedia.org/wiki/Semisimple_Lie_algebra}

\sbs{Levi decomposition and Heisenberg Lie algebra $\h_n$}
The symmetry algebra is a differential operator realization of a
semidirect sum of a semisimple algebra $S$ isomorphic to $\sl(2,\mathbb{R})\oplus \so(1,1)$
and a \emph{solvable ideal, i.e. the radical} $ R:= \h_2$, where  $\h_2$ denotes the Heisenberg algebra in 2 dimensions.
In the semidirect sum $S \semiplus R$, known as a \emph{Levi decomposition}, $R$ is an ideal of the Lie algebra
$S \semiplus R$, with
\be
[S,S]=S, \quad [S,R] \subseteq R, \quad [R,R] \subset R.
\ee
The Heisenberg Lie algebra $\h_n$ in $n$ dimensions (for $n = 1,2,\dots$) is
a $(2n+1)$--dimensional Lie algebra with generators $a_i$, $i = 1,\dots,n$,
and $a_i^+$, $i = 1,\dots,n$, and the unit operator 1, satisfying the commutation
relations
\be
[a_i,a_j] = 0, \quad [a_i^+,a_j^+] = 0, \quad [a_i,a_j^+] = \delta_{ij}, \qquad i,j=1,\ldots,n\,.
\ee

\sbs{Identifying $\spr{A_5,\ldots,A_9}$ with $\h_2$}
Let $J = A_4$, $D_1^+ = A_5$, $D_1^- = A_7$, $D_2^+ = A_8$, and $D_2^- = A_6$, then
\bes \lb{commJD}
\begin{align}
[ K_0,D_i^\pm ] &= \pm D_i^\pm, \quad i = 1,2, \lb{comm1} \\
{[} K_-,D_i^+ ] &= D_i^-, ~\quad i = 1,2, \lb{comm2} \\
{[} K_+,D_i^- {]} &= - D_i^+, \quad i = 1,2, \lb{comm3} \\[2mm]
{[} J,K_0 {]} &= [ J,K_- ] = [ J,K_+ ] = 0, \lb{comm4} \\[2mm]
{[} J,D_1^\pm {]} &= - D_2^\pm, \quad [ J,D_2^\pm ] = - D_1^\pm, \lb{jd}\\
{[} D_i^+,D_j^+ {]} &= [ D_i^-,D_j^- ] = 0, \quad i,j = 1,2, \lb{dd1}\\
{[} D_1^-,D_1^+ {]} &=  [ D_2^-,D_2^+ ] = 2, \lb{dd}
\end{align}
\ees
along with the commutators of the algebra $\sl(2,\mathbb{R})$, and all other commutators are 0.
In particular, $(D_i^+,D_i^-)$ is an $\sl(2,\mathbb{R})$ vector operator for both $i = 1$
and $i = 2$.
By \rf{dd1} and \rf{dd}, $D_1^\pm$, $D_2^\pm$, and $1$ generate a Lie algebra
isomorphic to $\h_2$ (e.g., set $a_1 = D_1^-$, $a_2 = D_2^-$,
$a_1^+ = \half D_1^+$, and $a_2^+ = - \half D_2^+$).
The element $J$ generates $\so(1,1)$.
Specifically, an orthogonal Lie algebra generated by a nondegenerate symmetric
bilinear form $g_{ij}$ on $\R^n$  has generators $J_{ij} = -J_{ji}$
for $i,j = 1,\dots,n$, satisfying commutation relations
\be
[J_{ij},J_{kl}] = g_{jk} J_{il} - g_{jl} J_{ik} - g_{ik} J_{jl} + g_{il} J_{jk}.
\el{ortho_rel}
The \emph{vector module} has a basis $<v_i>$, $i = 1,\dots,n$, and the action of
the generators on this module is given by
\be
J_{ij} v_k = g_{jk} v_i - g_{ik} v_j.
\el{ortho_vec}
A set of operators $v_i$, $i = 1,\dots,n$, is called a vector operator if
\be
[J_{ij},v_k] = g_{jk} v_i - g_{ik} v_j.
\el{ortho_vec_op}
Set the bilinear form on ${\R}^2$ so that $g$ is diagonal,
$g_{11} = +1$ and $g_{22} = -1$, and set
$J = J_{12}$, then the pairs $(D_1^+,D_2^+)$ and $(D_1^-,D_2^-)$ form
$\so(1,1)$ vector operators, so that $J$, $D_i^\pm$, and $1$ form a Lie
algebra isomorphic to $\so(1,1) \semiplus \h_2$.

\sbs{Contraction of $\so(3,1)$ to $\so(1,1) \semiplus \h_2$}

The Lie algebra is also a contraction of $\so(3,1)$.
Set the bilinear form on ${\R}^4$ so that $g$ is diagonal,
$g_{11} = +1$ and $g_{22} = g_{33} = g_{44} = -1$, and let $\ga $ be
a real number.
Set
\br
J          &=& J_{12}, \crn
\tilde D_i^+ &=& \ga  J_{i3}, \ i = 1,2, \crn
\tilde D_i^- &=& \ga  J_{i4}, \ i = 1,2, \crn
 C         &=& \ga ^2 J_{34}, \nn
\er
then
\br
[ J,\tilde D_1^\pm ] &=& - \tilde D_2^\pm, \crn
[ J,\tilde D_2^\pm ] &=& - \tilde D_1^\pm, \crn
[ \tilde D_1^+,\tilde D_2^- ] &=& [ \tilde D_1^-,\tilde D_2^+ ] = 0, \crn
[ \tilde D_1^-,\tilde D_1^+ ] &=&  [ \tilde D_2^+,\tilde D_2^- ] =  C, \crn
[ \tilde D_1^+,\tilde D_2^+ ] &=& [ \tilde D_1^-,\tilde D_2^- ] = \ga ^2 J, \crn
[ C,\tilde D_i^+ ] &=& \ga ^2 \tilde D_i^-, \quad i = 1,2, \crn
[ C,\tilde D_i^- ] &=& - \ga ^2 \tilde D_i^+, \nn
\er
and $J$ and $C$ commute with each other.
In the limit as $\ga  \to 0$ only the following commutators are non-zero:
\br
[ J,\tilde D_1^\pm ] &=& - \tilde D_2^\pm, \crn
[ J,\tilde D_2^\pm ] &=& - \tilde D_1^\pm, \crn
[ \tilde D_1^-,\tilde D_1^+ ] &=&  -[\tilde D_2^-,\tilde D_2^+ ] =  C, \nn
\er
and all other commutators are zero.
Identifying $\tilde D_i^\pm$ with $D_i^\pm$ and $C$ with $2$ (i.e. $2$ times
the identity operator) yields \rf{jd}, \rf{dd1} and \rf{dd}, and hence
the Lie algebra $ \so(1,1)\semiplus \h_2$, given above, is a \textbf{contraction} of $\so(3,1)$.

\be
\ba{|c|c|c|c|c|}
\hline
    & 1  &      2      &          3         &         4         \\  [1.2mm] \hline
%---------------------------------------------------------------------------				
1   &\quad 0 \quad  & J = J_{12}  & \tilde D_1^+ = \gamma J_{13} & \tilde D_1^- = \gamma J_{14} \\ [1.2mm]  \hline
2   &               &  0          & \tilde D_2^- = \gamma J_{23} & \tilde D_2^+ = \gamma J_{24} \\ [1.2mm] \hline
3   &               &             &      0                      & C = \gamma^2 J_{34}            \\ [1.2mm]  \hline
4   &               &             &                             & 0                           \\ [1.2mm]  \hline
\ea
\el{ContrD}
\begin{center}
Contraction diagram of $\so(3,1)$ to  $(\so(1,1) \semiplus \h_2)$. Recall $ D_1^+ = A_5, D_2^+ = A_8, D_1^- = A_7, D_2^- = A_6$.
\end{center}

The full symmetry algebra is now
$\g := (\sl(2,\mathbb{R})\oplus \so(1,1)) \semiplus \h_2$, as we wrote in \rf{symalg}.
Also, recall that $J$ commutes with $K_-, K_0$ and $K_+$.
These four operators form a representation of the Lie algebra $\gl(2,\mathbb{R})$.
(Note that $J$ is actually the center, and will be represented by a scalar
multiple of the identity in any two dimensional representation of $\gl(2,\mathbb{R})$,
along with a two-dimensional non-unitary representation of $\sl(2,\mathbb{R})$: its matrix counterpart must be a multiplier.)

The full symmetry algebra is now
$\g := (\sl(2,\mathbb{R})\oplus \so(1,1)) \semiplus \h_2$, as we wrote in \rf{symalg}.
Also, recall that $J$ commutes with $K_-, K_0$ and $K_+$.
These four operators form a representation of the Lie algebra $\gl(2,\mathbb{R})$.
(Note that $J$ is actually the center, and will be represented by a scalar
multiple of the identity in any two dimensional representation of $\gl(2,\mathbb{R})$,
along with a two-dimensional non-unitary representation of $\sl(2,\mathbb{R})$: its matrix counterpart must be a multiplier.)

\subsection{Realization of Virasoro algebra in terms of $t^n L$}

By noting the commutator $[L,t]=1$, we can easily check that the three
operators $(L,tL,t^2L)$ have the following commutation relations
\be
[tL,L]=-L~, \qquad
[tL,t^2L]=t^2 L~, \qquad
[L,t^2L]= 2 tL~.
\el{suL}
and therefore yield a realization of the generators $<K_-,K_0,K_+>$ of
$\su(1,1)$.

It is interesting to note that if we define
$d_n\equiv
-t^{n+1}L$ for $n\in \Z$, then
$d_n$ satisfy the commutation relations of the Virasoro algebra
\be
[d_m,d_n]=(m-n)d_{m+n}+ \delta_{m,-n} \frac{m(m^2-1)}{12} c
\el{va}
with a zero central term $c=0$.

These commutation relations are not obvious, since the $d_n$
contains in addition to $\widetilde{d}_n\equiv -t^{n+1}\pdt$,
which is a well-known realization of the algebra \rf{va},
also the terms $(1/4) t^{n+1}(\pdx^2-t^{-2}\pdp^2)$, which depend on $t$.

\sbs{$\su(1,1) \semiplus\text{Virasoro}$}

By applying the commutation condition $ [L, A]=\xi(x,p,t) L$, from \rf{rq} and noting
that $\xi=2c_3t+2c_2$ from \rf{cs1c}, we obtain
$$
L A = L~\sum_{i=1}^9 c_i A_i = \xi= 2c_3 t+2c_2, $$
which yields
\be
[L,A_1]= 0, \qquad[L,A_2]= 2L, \qquad  [L,A_3]= 2tL.
\el{tla}
This tells us that all the $A_i$, except for $K_0=A_2$ and $K_+ =A_3$, commute with $L$.

By noting that $\spr{A_1, A_2, A_3}$ isomorphic to $\su(1,1) := \spr{K_-, K_0, K_+}$,  we obtain
\be
[ t^m L,K_+ ] = (2-m)~ t^{m+1} L, \quad [ t^m L,K_0] = (1-m) ~t^{m} L, \quad [L,K_- ] = -m ~t^{m-1}L,
\el{tlk}
which yields an infinite algebra, which we identify as follows:
$$ \su(1,1) \semiplus V_0 := \spr{K_-, K_0, K_+} \semiplus \spr{t^m L , m\in \mathbb{Z} },$$
because $V_0$ is an ideal.

%==================================================================================

\end{appendices}

\edd